\documentstyle[12pt,aaspp4]{article}

\begin{document}


\def\zz{\hang\noindent}

\def\kms{km s$^{-1}$}
\def\pix{pix$^{-1}$}
\def\deg{$^\circ$}

\def\mic{{$\mu$m}}

\def\h2o{H$_2$O}

\def\ak{{\it $A_K$}}

\def\teff{$T_{\rm eff}$}

\def\aple{$\mathrel{\hbox{\rlap{\hbox{\lower4pt\hbox{$\sim$}}}\hbox{$<$}}}$
}
\def\apge{$\mathrel{\hbox{\rlap{\hbox{\lower4pt\hbox{$\sim$}}}\hbox{$>$}}}$
}



\title{The Stellar Content of Obscured Galactic Giant\footnote{
In this series of papers we shall follow a suggestion of Dr.
Robert Kennicutt (private communication) that ``giant'' means that more
than $10^{50}$ Lyman continuum photons are inferred to be emitted per
second from the H~II region. This is about ten times the luminosity of the
Orion nebula and roughly the number emitted from the hottest {\it single}
O3-type star.  As these stars are not found in isolation, there is an
implication that a ``giant'' H~II region contains some minimum of {\it
multiple} O-type stars.} H~II Regions: I. W43}

\author{R. D. Blum\altaffilmark{2}} 
\affil{Cerro Tololo Interamerican Observatory, Casilla 603, La Serena,
Chile\\ rblum@noao.edu}

\author{A. Damineli\altaffilmark{2,3}}
\affil{ IAG-USP, Av. Miguel Stefano 4200, 04301-904, Sao Paulo, Brazil
\\damineli@iagusp.usp.br}

\author{P. S. Conti}
\affil{JILA, University of Colorado\\Campus Box 440, Boulder, CO,
80309\\pconti@casa.colorado.edu}

\altaffiltext{2}{Visiting Astronomer, Cerro Tololo Interamerican
Observatory,
National Optical Astronomy Observatories, 
which is operated by Associated Universities for Research in Astronomy,
Inc., 
under cooperative agreement with the National Science Foundation}

\centerline{{\it accepted for publication in the} Astronomical
Journal} 

\begin{abstract}

Near infrared images of the Galactic giant H~II region W43 reveal a dense
stellar cluster at its center. Broad band 
$JHK$ photometry of the young cluster and
$K-$band spectra of three of its bright stars are presented. The 2 \mic \
spectrum of the brightest star in the cluster is very well matched to the
spectra of Wolf--Rayet stars of sub--type WN7. Two other stars are
identified as O type giants or supergiants by their \ion{N}{3} and \ion{C}{4}
emission. The close spatial clustering of O and the hydrogen WN type stars
is analogous to the intense star burst clusters R136 in the Large
Magellanic Cloud and NGC~3603 in the Galaxy.
 
\end{abstract}


\newpage

\section{INTRODUCTION}

Near infrared (1--5 \mic) spectroscopic classification schemes have
recently been outlined for OB stars (Hanson et al. 1996\markcite{hcr96},
hereafter HCR96; Blum et al. 1997\markcite{beal97}) and Wolf--Rayet (WR)
stars (e.g., Eenens et al. 1991\markcite{eww91}; Figer et al. 
1997\markcite{fmnl97}). Coupled with infrared spectrometers on large
telescopes, these classification schemes are now pushing forward the
exploration of optically obscured, young stellar populations throughout the
inner Galaxy. Indeed, Hanson et al. (1997)\markcite{hhc} have presented
a detailed investigation of the ionizing O and B stars in M17, a relatively
nearby but partially obscured giant 
HII region. In addition, the central 50 pc of the Milky Way have been
intensely studied with the application of near infrared spectroscopy. It
has been revealed that the stellar cluster in the central parsec, as well
as two other nearby clusters, are rich in OB and Wolf--Rayet stars (see
Morris \& Serabyn 1996\markcite{ms96} for a recent review and the many
references therein).

Massey et al. (1995)\markcite{mjd95} have produced a detailed
characterization of the stellar mass function in optically visible OB
associations in the Milky Way. Their fundamental work serves as a benchmark
comparison of Galactic OB star properties relative to other environments
(e.g., the Magellanic Clouds). The present project, in which this paper is
a beginning, will build on Massey et al.'s work by investigating the 
optically obscured stellar content of the inner Galaxy giant H~II regions
through infrared photometry and spectroscopy. These observations will probe
and compare the mass function as a function of Galacto--centric position,
provide checks on Galactic structure (spiral arm morphology) through
spectroscopic distance estimates which can be compared to radio
recombination line/rotation model distances, and investigate massive star
formation phenomena. The latter aspect is particularly exciting as many of
the optically obscured giant H~II regions are young enough that we expect
to find multiple epochs of star formation, i.e., young stellar objects
(YSOs) in the presence of more evolved OB and WR stars. This is already in
evidence in M~17 where it appears that while the hottest O stars have
little or no infrared excess (having ``blown'' away their birth cocoons), 
the somewhat less massive later type O and B stars show evidence for discs
still surrounding them (Hanson et al. 1997\markcite{hhc}). These two groups
are spatially segregated in M~17. Are the early O stars more efficient at
removing their circumstellar material through strong winds and radiation,
or do the two groups represent star formation at different epochs in M~17?
Investigations of other giant H~II regions will help answer such
fundamental questions about massive star formation.

W43 (=G30.8-0.2) is a giant H~II region in the first Galactic quadrant
($l, b = 30.8, -0.2$).
Its distance, from H~I line radio measurements and a Galactic rotation
model, is about 7 kpc (Smith et al. 1978\markcite{sbm78}). The entire
region emits about $10^{51}$ Lyman continuum photons per sec (= Lyc)
according to these authors. This is equivalent to roughly 100 ``O7 star
equivalents'' (Vacca 1994\markcite{v94}), similar to NGC~3603 and a factor
of a few smaller than R136. W43 contains about $10^6$ solar masses of
molecular gas, a quantity which suggests ongoing star formation (Liszt
1995\markcite{l95}). W43 has been 
observed by Liszt et al. (1993)\markcite{lbg93}
with the
Very Large Array (VLA): 
they provide a 9$'\times9'$ scale map in the radio L band continuum
synthesized from the H~I absorption measures. The overall spatial morphology
of the emission is roughly that of a ``T''with a sagging ``wing'' to the
west.  Subrahmanyan \& Goss (1996)\markcite{sg96} present an 
80$'\times80'$  VLA map
of the entire W43 complex at 330 MHz; a number of other sources are
indicated, but G30.8-0.2 dominates the radio emission.  Liszt et al.
(1993)\markcite{lbg93} call the position of the strongest source G30.78--0.23; 
this is 
the top bar of the ``T''. This source is called G30.8N in the far infrared
maps of  Lester et al. (1985)\markcite{ldw85}. 
The bottom of the ``T'', also seen in the 
far infrared maps, is denoted G30.8S by Lester et al. 
(1985)\markcite{ldw85}.
The stellar cluster, which will be the main topic
of this paper, lies between G30.8N and G30.8S and was first seen
as an unresolved $K-$band source by Lester et al. (1985)\markcite{ldw85}. 

Using $J$, $H$, and $K-$band photometry, we reveal this stellar source
as a massive cluster of newly formed stars. 
Three of the most luminous
stars in the cluster are identified and classified as hot stars.
W43 is totally
obscured optically; we estimate (below) that $A_{V}$ is about 30 mag.
 
\section{OBSERVATIONS AND DATA REDUCTION}

High angular resolution $J$, $H$, and $K$ images of W43  were
obtained on the nights of 28 and 29 August 1998 with 
the f/14 tip--tilt system on the Cerro Tololo Interamerican Observatory
(CTIO) 4m Blanco telescope using the facility
imager (CIRIM). 
The spectroscopic data were
obtained on the night of 02 June 1998 using the f/14 tip--tilt system at
the Blanco telescope with the facility infrared spectrometer (IRS).
Spectroscopic 
pointings were made from preliminary images taken with the Cryogenic
Optical Bench (COB) obtained on 24 April 1997.
CIRIM is described in the instrument manual found on the CTIO www pages
(www.ctio.noao.edu). The tip-tilt system, COB imager, and IRS are described
on the CTIO www pages as well as by P\'{e}rez \& Elston
(1998)\markcite{pe98}, Probst et al. (1994)\markcite{peal94}, and DePoy et
al. (1990)\markcite{deal90}, respectively. Briefly, the tip--tilt system
uses three piezo--electric actuators to move the secondary mirror at high
frequency in a computer controlled feed--back loop to compensate for 
image centroid motion. COB employs 0.094$''$ pixels, the IRS 0.32$''$
pixels, and CIRIM 0.21$''$ pixels.

All basic data reduction was accomplished using
IRAF\setcounter{footnote}{2}\footnote{IRAF is distributed by the
National Optical Astronomy Observatories.}. Each image/spectrum 
was flat--fielded using dome flats and then sky subtracted using
a median combined image of five to six frames. For W43 itself,
independent sky frames were obtained five arcminutes east and/or south
of the cluster. Standard stars used the median combination of the data for
sky.

\subsection{Images}

The CIRIM image data were obtained under
photometric conditions and in 0.50$''$ \aple FWHM \aple 0.80$''$
seeing (this includes the tip--tilt correction). Total exposure times
were 720~s, 90~s, and 90~s at $J$, $H$, and $K$, respectively.
The individual $J$, $H$, and $K$ frames were shifted and combined. DAOPHOT
(Stetson 1987\markcite{s87})
photometry was performed on the combined images using brighter stars off
the cluster center for the psf reference (The LickVista 5.0 version
of DAOPHOT was used). The flux calibration was
accomplished using standards 
9172 and
9178 from Persson et al. (1998)\markcite{peal98} which are essentially on
the CIT/CTIO photometric system (see the discussion on color
transformations in Persson et al. 1998\markcite{peal98}). 
The standards were taken just before and after the
W43 data, bracketing the airmass of W43, except for the $K-$band
images for which the standard and W43 were within 0.1 airmass of each other.
Aperture corrections using 8
pixel radius apertures were used to put the instrumental magnitudes on the
CTIO/CIT system. The central star in the cluster was saturated in the CIRIM
$K-$band images. Its magnitude has been corrected (by 0.06 mag) using the 
preliminary COB data taken in 1997.
The $J-$band data include
a color correction as given in the CIRIM manual ($-$0.037 $\times$ $J-K$).

Uncertainties for the final $JHK$ magnitudes include the formal DAOPHOT error
added in quadrature to the error in the mean of the photometric standards.
The latter was derived from the scatter in the individual measurements and
was $\pm$ 0.027, $\pm$ 0.022, $\pm$ 0.013 mag in $J$, $H$, and $K$, 
respectively. The DAOPHOT errors ranged from approximately 
$\pm$ 0.01 mag to an arbitrary cut--off of 0.5 mag 
(stars with larger errors were excluded from further analysis).

Lower angular resolution images were obtained at $J$, $H$, and $K$ using 
CIRIM at f/13.5 on the CTIO 1.5m (0.65$''$ pix$^{-1}$).
The individual $J$, $H$, and $K$ frames
were shifted and combined and 
have measured seeing of $\approx$ 1.5$''$ FWHM.
While these low resolution images are calibrated, their photometry is
not discussed further in this paper. They are used only to discuss 
the morphology of the region surrounding the high resolution images of the 
W43 cluster.

\subsection{Spectra}

The spectra were obtained with a 0.65$''$ wide slit (oriented NS) in \aple
1$''$ seeing  and divided by the spectrum of HD7209 (A1V) to remove
telluric absorption features. Br$\gamma$ absorption in HD7209  was removed
by eye by drawing a line across it between two continuum points. One
dimensional spectra were obtained by extracting and summing the flux in a
$\pm$ 2 pixel aperture.

The wavelength calibration was accomplished by measuring the positions of
six bright OH$^-$ lines from the $K-$band sky spectrum (Oliva \& Origlia
1992\markcite{oo92}). Because the grating was moved when acquiring objects,
the zero point of the solution is good only to a few pixels; lines are
identified by their relative difference between one and another. The
measured dispersion is 0.001323 \mic \ pix$^{-1}$. The spectral resolution
at 2.2 \mic \ is $\lambda/\Delta\lambda \approx$ 800. 

\section{RESULTS}

\subsection{Images}

\begin{figure}
\plotfiddle{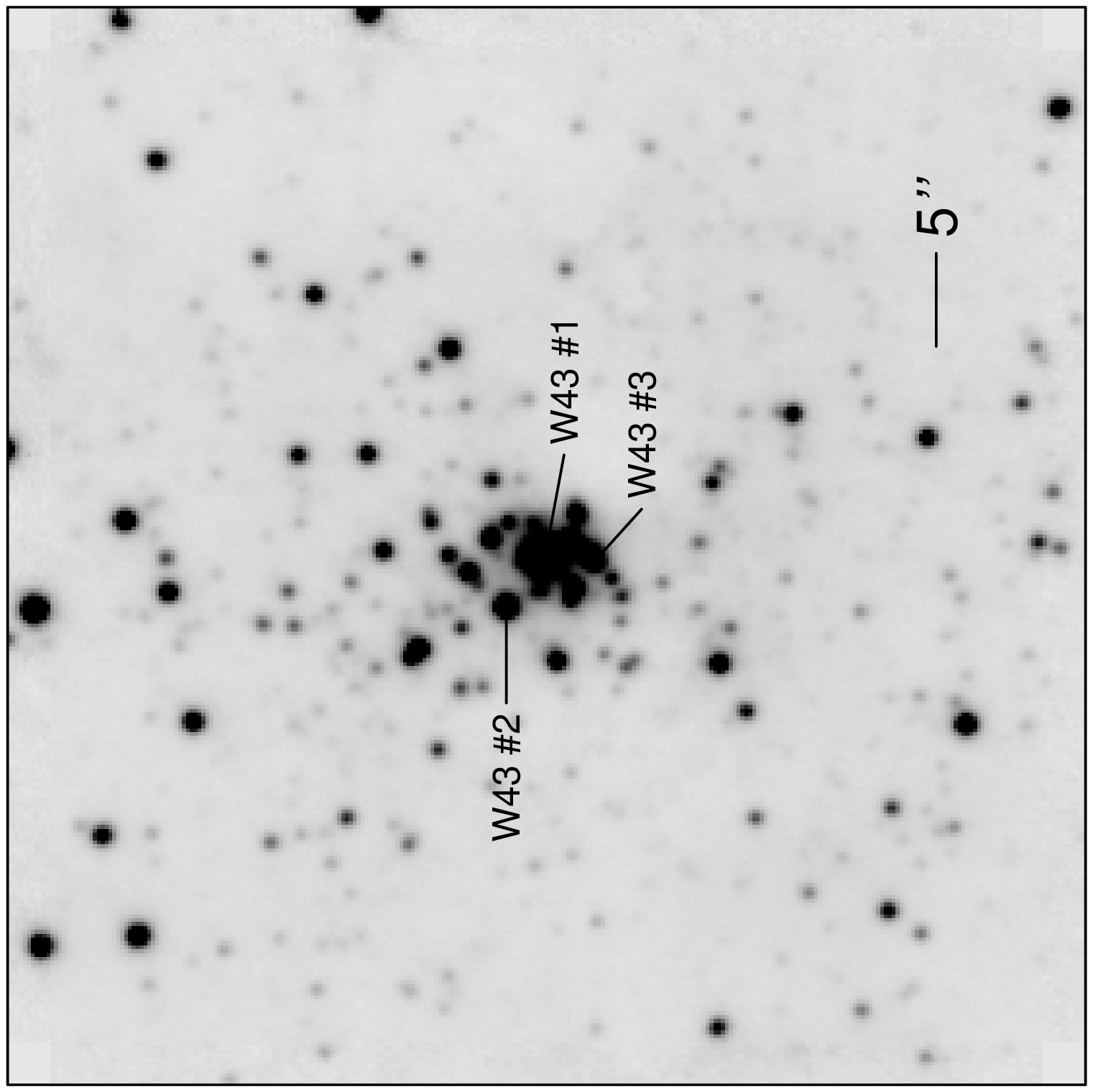}{7.0 in}{-90}{45}{45}{-180}{590} 
\plotfiddle{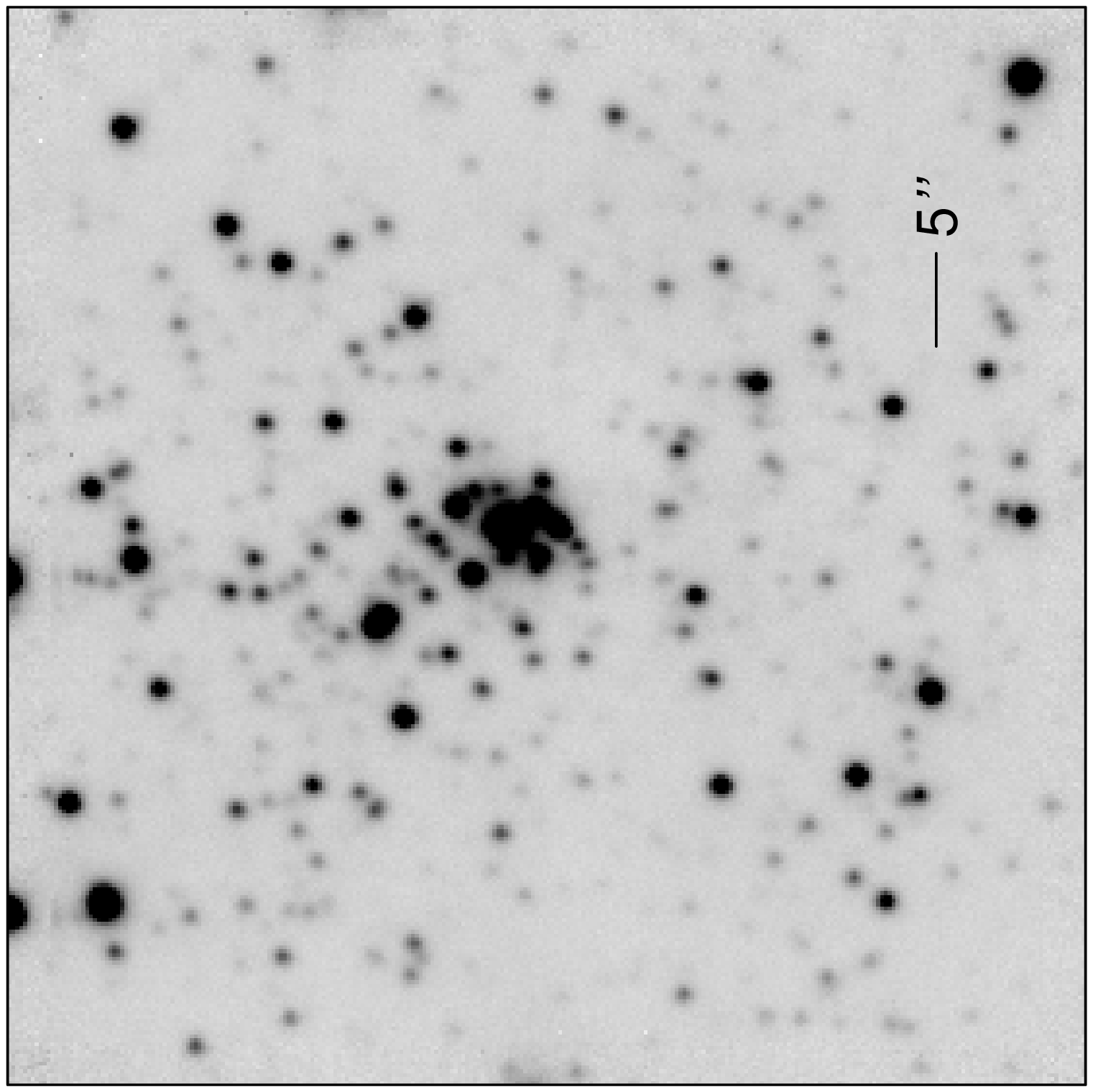}{.1  in}{-90}{45}{45}{-180}{415} 
\plotfiddle{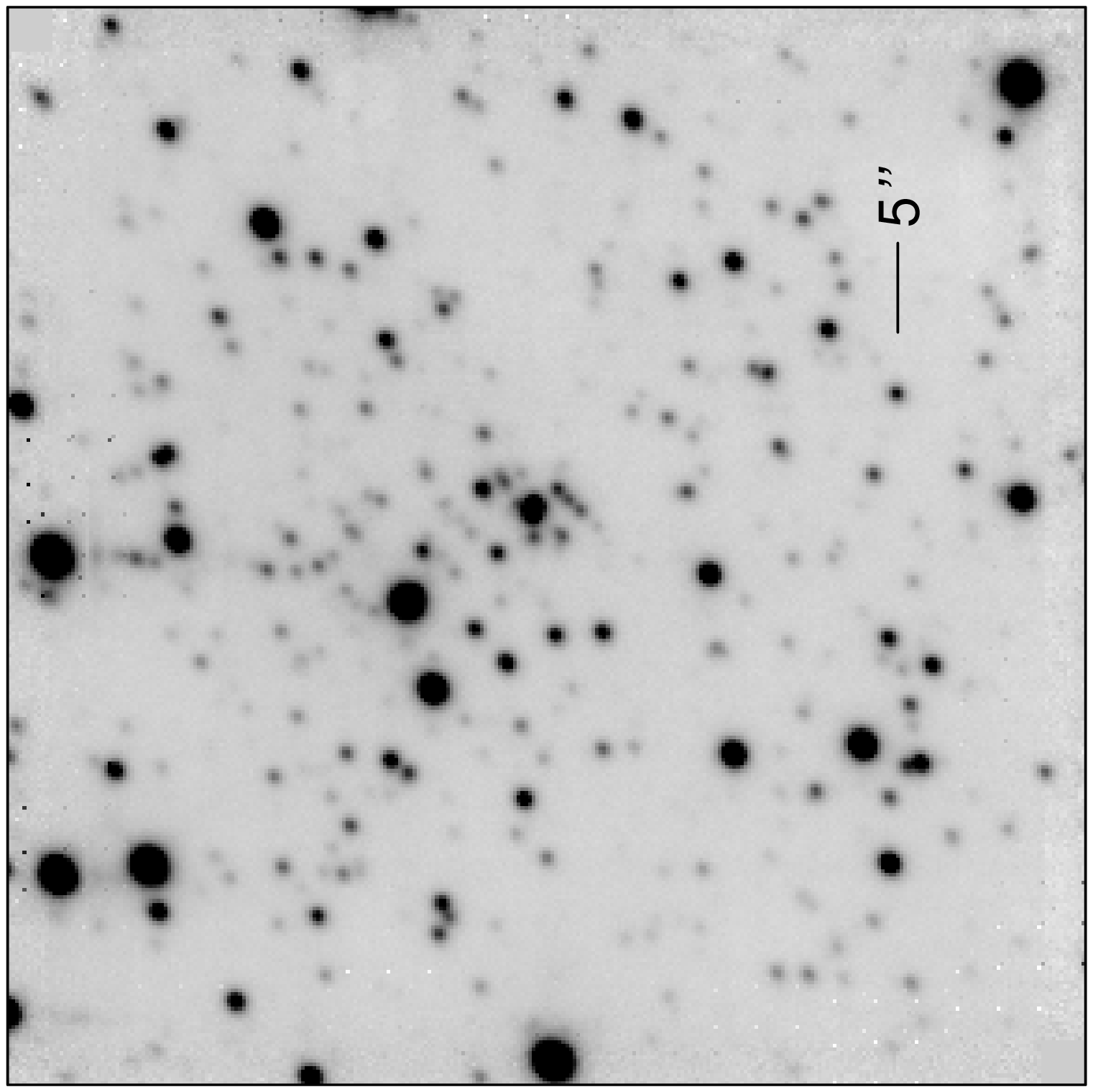}{.1  in}{-90}{45}{45}{-180}{240} 
\figcaption[]{
High resolution CIRIM images of the W43 central cluster. $K$ (2.2 \mic)
is at the top, $H$ (1.65 \mic) in the middle, and $J$ (1.25 \mic) at
the bottom. North is up, East to the left.
Three of the brightest four stars in the central $\sim$ 5$''$ have spectra
presented in Figures~\ref{w431}, \ref{w432}, and \ref{w433}.
\label{w43k}
}
\end{figure}

$J$, $H$, and $K$ high angular resolution images are shown in
Figure~\ref{w43k}. The presence of a highly reddened cluster is
immediately apparent. An unresolved $K-$band source was reported at this
position by Lester et al. (1985\markcite{ldw85}). 
The source was not centered 
on the far infrared peaks in their maps, 
but was centered on a peak in a 6 cm radio map which they
also presented. We discuss the position of the cluster relative to the 
far infrared emission and radio sources in \S3.1.2.
Cotera \& Simpson (1997)\markcite{cs97} also
noted the presence of a stellar cluster from narrow-band images at 2 \mic \
(see \S3.2), probably at this location.

\subsubsection{Color--Magnitude and Color--Color Plots}

\begin{figure}
\plotone{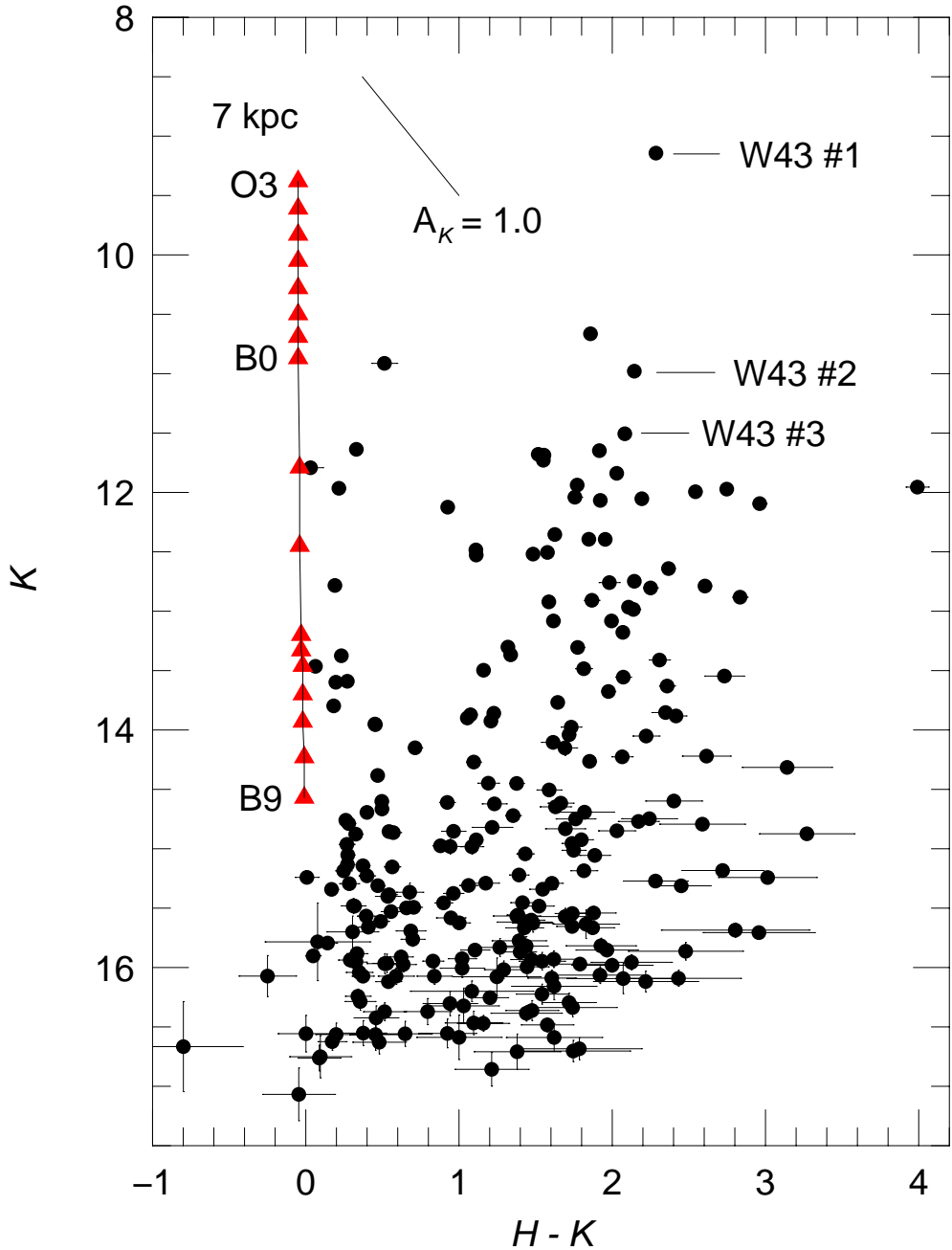}
\figcaption[]{$H-K$ color--magnitude diagram (CMD) for the W43 cluster. 
Data are from the high angular resolution CIRIM images. A 
foreground sequence is clearly present as well as a cluster sequence and a 
number of very red stars. Stars for which spectra are presented are
labeled. The main sequence is plotted as {\it filled} triangles for
spectral types O3 to B9 at a distance of 7 kpc.
\label{cmd}
}
\end{figure}

The $K$ vs. $H-K$ color--magnitude diagram (CMD) is
given in Figure~\ref{cmd}. This diagram is derived from a roughly 58$''$
$\times$ 58$''$ area around the central cluster.
The CMD shows a cluster sequence near $H-K$ 1.8 $\pm$ 0.4 mag. The observed
scatter is attributed to differential reddening (see below). Stars W43~\#1,
W43~\#2, and W43~\#3, for which spectra are presented, are labeled on the 
diagram. The main sequence is also shown in Figure~\ref{cmd} for O3 to B9 stars
(Vacca et al. 1996\markcite{vgs96}, O3--B0
; Schmidt-Kaler 1982\markcite{sk82}, B1--B9) assuming a distance of 7 kpc 
(\S 1).

Figure~\ref{cc} shows a $J-H$ vs. $H-K$ color--color plot of the stars on
the high resolution 
CIRIM images. This diagram indicates a large range in extinction and
reddening in the field (0 \aple \ak \aple 4.5 mag for stars detected at
$J$, $H$, and $K$).
The {\it dashed} lines in this plot represent
reddening lines due to interstellar extinction based on the average
relation given in Mathis (1990\markcite{mat90}). Each of the two 
lines begins at the colors for an unreddened star (M3 III, Frogel et al.
1978\markcite{fpam}; O6, Koornneef 1983\markcite{k83}). The lines describe
an envelope of reddened colors for a nearby ``disk'' population with
the above mean colors. 
From Figure~\ref{cmd}, it is clear that the 
cluster sequence lies near $H-K$ $\sim$ 2 mag. This corresponds to
\ak \ of roughly 3.2 mag (or $A_V$ $\approx$ 34 mag).
Estimates of the \ak \ toward the spectroscopic targets are made
in \S3.2 where the spectroscopic clarification is used to determine
the intrinsic colors of the stars.

Several objects on the color--color 
plot have a large excess emission indicated.
Such excesses are consistent with the objects being YSOs with 
circumstellar material (e.g., a disk as 
indicated for a number of newly formed 
stars in M~17 -- Hanson et al. 1997\markcite{hhc}, though we have no evidence
yet for any particular geometry giving rise to the putative excess emission
in this case). 
There are also objects which appear extremely red in Figure~\ref{cmd} but
which are not detected in $J$. These may be seen through even larger
columns of obscuring material.  Alternately, they may be intrinsically 
fainter at $J$ but have a larger excess at $K$. 
These possibilities
will be investigated with future spectroscopic observations.

\subsubsection{Wide Field of View Image}

\begin{figure}
\plotone{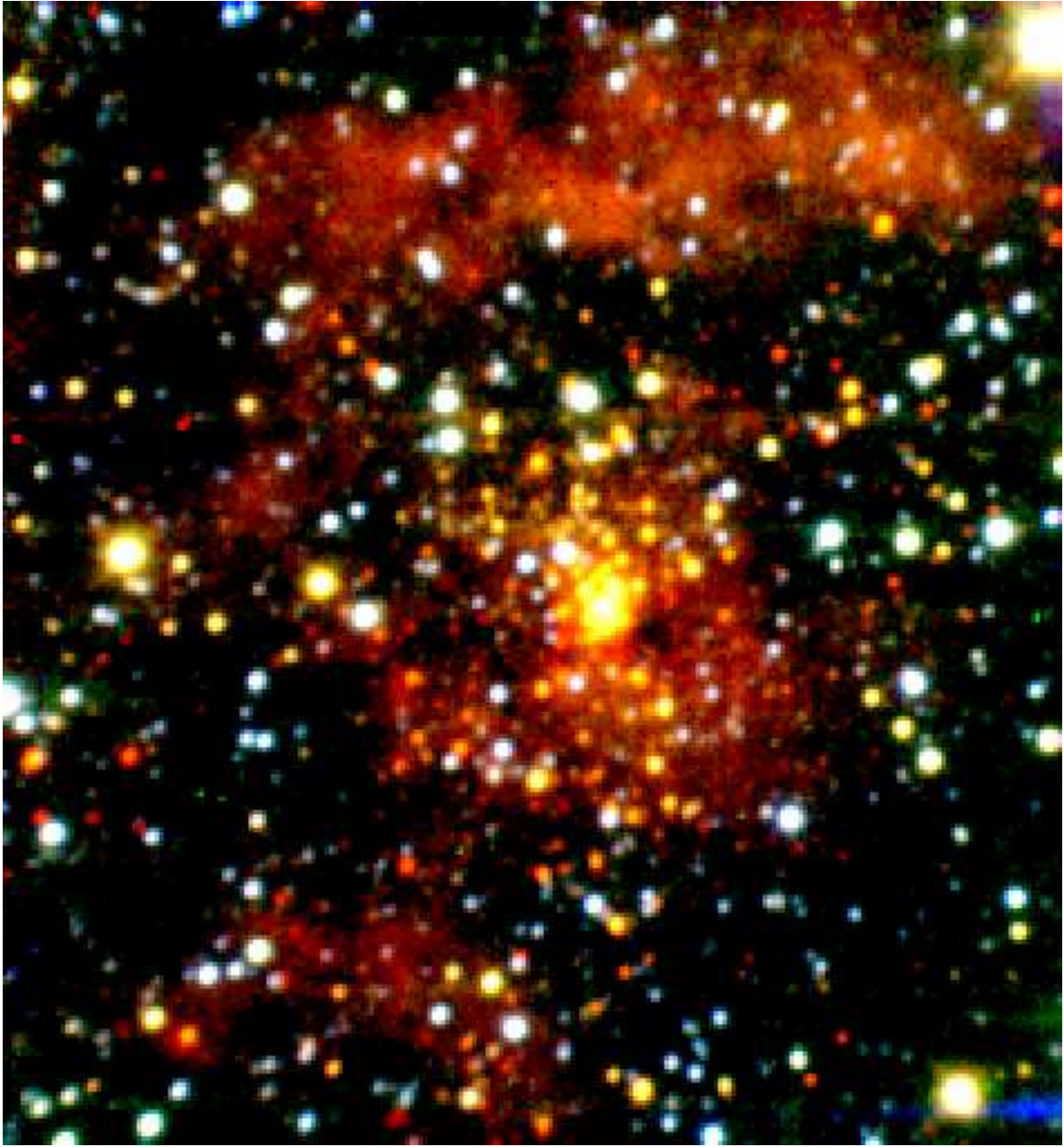}
\figcaption[]{
$JHK$ color composite image of the W43 cluster and surrounding field made
from the 0.65$''$ pix$^{-1}$ CIRIM images
($J$ is coded blue, $H$ green, and $K$ red). North is up, East to the left.
The field of view is approximately 2.7$'$ $\times$ 2.7$'$. 
There is strong differential reddening in the field (see text) with darker
clumps of material quite close in projection to the central cluster 
(e.g $\sim$ 5$''$ SW). The E--W nebulosity at the top is the region
G30.8N in the terminology of Lester et al. (1985); the nebulosity to the
SE of the cluster is G30.8S.
\label{clr2}
}
\end{figure}

A $JHK$ color composite of the wide field of view CIRIM images 
(2.7$'\times2.7'$) is shown in Figure~\ref{clr2}.
This composite demonstrates that much of the scatter in Figure~\ref{cmd}
(especially in the cluster sequence)
can be attributed to 
differential reddening. The spatial scales 
involved can be quite small.
For example, there is  
an obvious clump of obscuring material 
which appears in projection just arcseconds to the SW of the  central
cluster.

Along the upper edge of Figure~\ref{clr2} nebulosity (from Br$\gamma$) is
present. This is the upper part of the ``T'' so obvious in the radio and
far infrared images (Lester et al. 1985\markcite{ldw85}; 
Liszt et al. 1993\markcite{lbg93}), and denoted by Lester et al. as
G30.8N.  
Very red stars appear in projection against G30.8N in
the color image.
Notice that in the $JHK$ color image there is {\it no} obvious vertical
part of the ``T''. However, there is nebulosity and (again) a clustering of red
stars to the southeast of the cluster. This is part of the far infrared source
which Lester et al. (1985\markcite{ldw85}) call G30.8S.
According
to Lester et al. (1985\markcite{ldw85}) there are two radio sources 
in between G30.8N and G30.8S which lie in a pronounced minimum
of the far infrared emission. The western one of these was
described by them as an unresolved $K-$band 
source ($\sim$ 8th magnitude) which they claimed was the source
of ionization for G30.8. This claim is confirmed in great detail by the 
images and spectra presented here.
We have confirmed the location of the Lester et al. $K-$band source as the
stellar cluster by comparing our near infrared images to a 
Digitized Sky Survey
\footnote{
Based on photographic data obtained using The UK Schmidt Telescope.     
The UK Schmidt Telescope was operated by the Royal Observatory          
Edinburgh, with funding from the UK Science and Engineering Research    
Council, until 1988 June, and thereafter by the Anglo-Australian        
Observatory.  Original plate material is copyright (c) the Royal        
Observatory Edinburgh and the Anglo-Australian Observatory.  The        
plates were processed into the present compressed digital form with     
their permission. The Digitized Sky Survey was produced at the Space   
Telescope Science Institute under US Government grant NAG W-2166.}
image centered on their coordinates 
($\alpha_{\rm 1950} = 
18^{\rm h} \ 45^{\rm m} \ 00^{\rm s}, \delta_{\rm 1950} = -01^{\circ} \
59' \ 54''$).

\subsection{Spectra}

\begin{figure}
\plotone{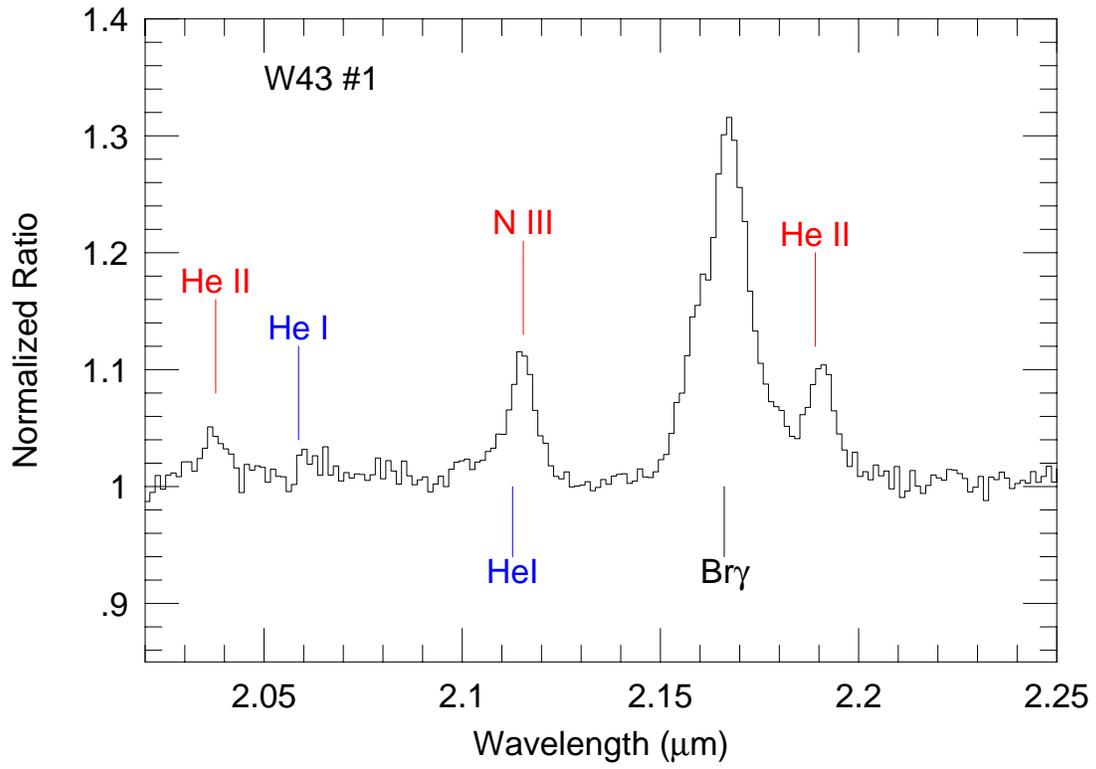}
\figcaption[]{W43~\#1, the brightest star 
of the W43 cluster. Comparison to similar spectra
of known  Wolf--Rayet (WR) stars in Figer et al. (1997) 
indicates this object is a member of the nitrogen sequence (WN7). 
\label{w431}
}
\end{figure}

\begin{figure}
\plotone{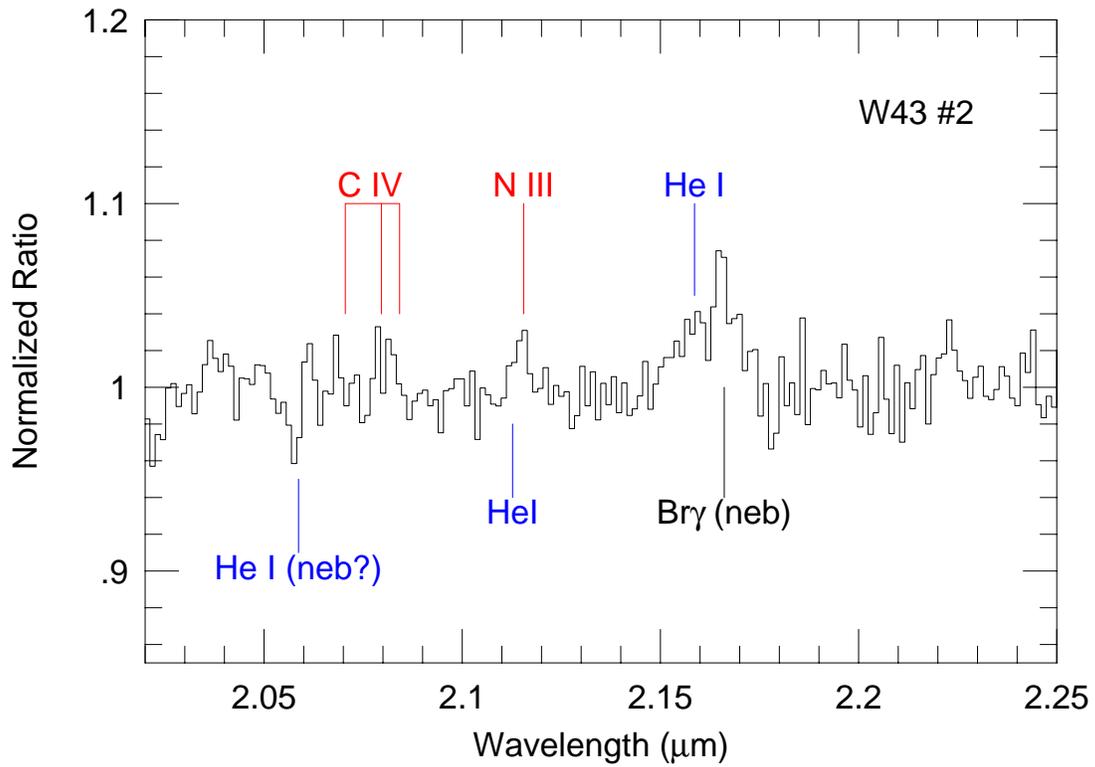}
\figcaption[]{W43~\#2. Comparison to similar spectra of known hot stars from
Hanson et al. (1996) suggests this is an early O type supergiant.
The Br$\gamma$ emission is probably nebular in origin; its contribution is
highly non-uniform along the slit.  Similarly, the \ion{He}{1} 2.06~\mic \  
line is possibly over--subtracted nebular emission. 
Positions of other He~I lines are indicated, though the lines are
not necessarily detected.
\label{w432}
}
\end{figure}

\begin{figure}
\plotone{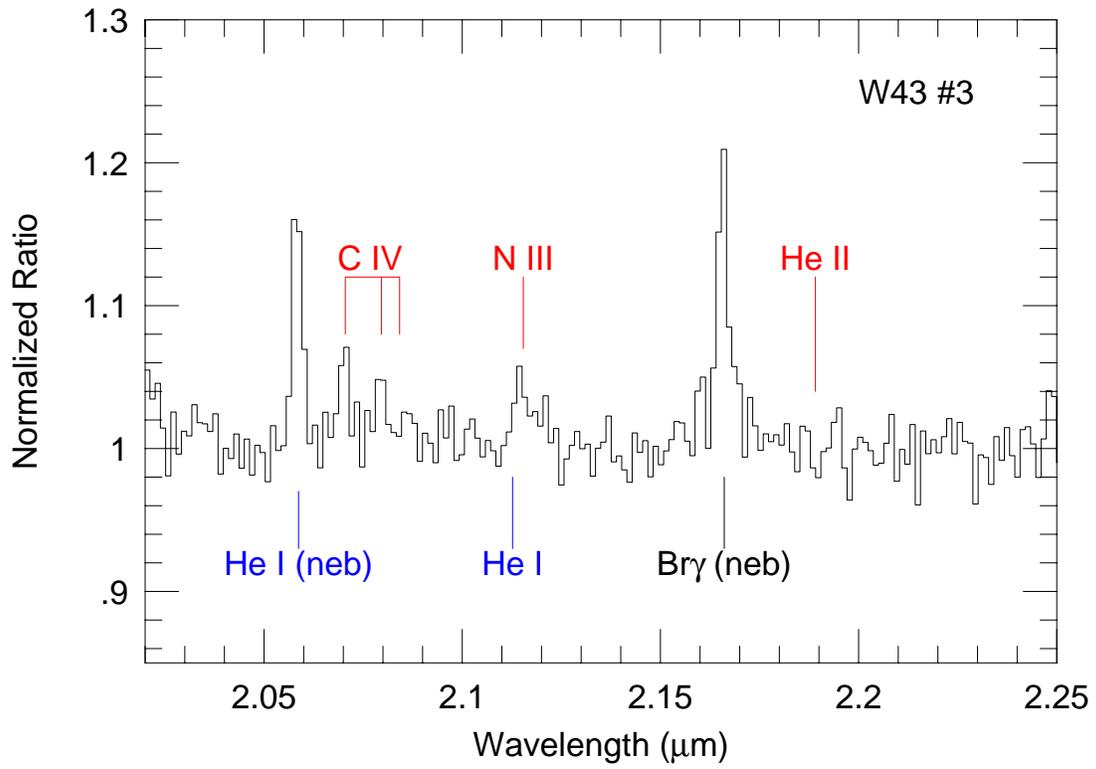}
\figcaption[]{
W43~\#3.
Comparison to similar spectra of known hot stars from 
Hanson et al. (1996) suggests this is an early O type supergiant.
The Br$\gamma$ and \ion{He}{1} 2.06~\mic \ features are probably nebular;
see text.
\label{w433}
}
\end{figure}

The spectroscopic targets, W43~\#1, \#2, and \#3 (Figures~\ref{w431},
\ref{w432},
\ref{w433}), were chosen from the preliminary 
COB CMD mentioned in \S2.1 
(similar but not as deep as Figure~\ref{cmd}) as the brightest
likely members of the central cluster. The second
brightest star is only 0.5$''$ E of W43~\#1 and was not observed 
spectroscopically because of
its proximity to W43~\#1. The effect of this star on the spectrum of
W43~\#1 should not be too great since it was placed outside the 0.65$''$ 
slit, has
similar color to W43~\#1, and is only 25$\%$ as bright as W43~\#1.

\subsubsection{W43~\#1}

The line widths and equivalents widths (EW) of W43~\#1 are given in
Table~\ref{tab1} and compare well with those in Figer et al.
(1997\markcite{fmn97}) for WR~131, a WN7 star of the hydrogen rich nitrogen
sequence. 
The strong \ion{N}{3} 2.1155~\mic \,  \ion{He}{2} 2.189~\mic \
and Br$\gamma$ $+$ \ion{He}{1} (near 2.17~\mic) emission  are the
signatures of a WN star. Cotera \& Simpson (1997)\markcite{cs97} report
\ion{He}{2} emission in narrow--band near infrared images of W43. This WN star
is probably the source of the emission they observed. 
There is some similarity among the $K-$band spectra of some Of stars and 
late type WN
stars (Conti et. al. 1995\markcite{ceal95}), but the excellent match to the
spectrum for WR131 for W43~\#1 and the large line widths ($\sim$ 1400 \kms)
argue strongly for a WN classification. Comparing the line EW's and line
widths to the other WN stars in Figer et al. (1997\markcite{fmn97})
indicates that W43~\#1 is not as hot as a WN6 type because these objects
show stronger \ion{He}{2} (2.19 \mic) emission relative to Br$\gamma$.
W43~\#1 does not appear as cool as later type WN8 and WN9 stars which
exhibit weaker \ion{He}{2} emission and/or much stronger \ion{He}{1} 
(2.06~\mic) emission.
The long slit image of W43~\#1 indicates no significant nebular contamination
to the stellar spectrum. Comparison to the line ratios in Figer et al.
(1997\markcite{fmn97}) also suggests this is the case.

As noted above, W43~\#1 has a $K-$band spectrum similar to the optically 
classified WN7 star WR~131 (Figer et al. 1997\markcite{fmn97}).
More precisely, WR~131 has an optical
spectral classification of WN7$+abs$, where $+abs$ refers to absorption
lines seen in the {\it optical} spectrum. The absorption has 
some times been attributed to an unresolved companion star.
While there is no evidence of absorption in the $K-$band spectrum
of W43~\#1 or the $K-$band spectrum of WR~131, 
it possible that W43~\#1 also has such a companion. This possibility
is suggested by the fact that
the standard star WN7 $K-$band 
spectra appear to show a relative dilution of the 
$K-$band features between those stars with and without {\it optical}
absorption lines (Figer et al. 1997\markcite{fmn97}, compare
WR120 and WR131; also P. Eenens, private communication). And, it is the
standards with the $+abs$ to which W43~\#1 is most similar.

\subsubsection{W43~\#2}

The spectrum of W43~\#2 is shown in Figure~\ref{w432}. 
\ion{N}{3} at 2.1155 \mic, is
clearly detected. The N~III emission suggests that
W43~\#2 is a mid to early O star (HCR96). 
In addition, there appears to be 
weak C~IV emission at 2.078 \mic \ (this line is typically less than 
one angstrom EW in somewhat higher resolution spectra of O stars, HCR) 
and perhaps some stellar Br$\gamma$
emission. 
The overall Br$\gamma$ feature is predominately nebular emission;
it is difficult to perfectly 
remove this contribution as it is not uniform across
the slit and this region of the spectrum was necessarily corrected with the
telluric standard. \ion{He}{2} 2.189~\mic \ absorption is not detected but
can be weak or in emission for
supergiant O stars (HCR). The absence of \ion{He}{2} 2.189~\mic \
absorption in the spectrum of W43~\#2 may be naturally explained if it is a
supergiant. A supergiant classification is consistent with its position in
the $K$ v. $H-K$ CMD (Figure~\ref{cmd}; see below). The \ion{He}{1} line at
2.06~\mic \ is in absorption but this is probably an over--subtracted nebular
contribution and/or imperfect telluric absorption correction.

\subsubsection{W43~\#3}

W43~\#3 has a $K-$band spectrum (Figure~\ref{w433}) which suggests it too is
an O star. \ion{N}{3} is detected as well as \ion{C}{4} emission.
The 2.078 \mic \ line appears slightly weak relative to the 2.069 \mic \ line
(the former is typically stronger; see HCR). This may not be too much of 
a concern in light of the weakness of the lines and the apparent 
signal to noise.
Strong emission features at \ion{He}{1} 2.06~\mic \ and Br$\gamma$ are also
present. \ion{He}{1} 2.06~\mic \ is not expected in normal O star spectra 
(HCR96) so this emission (and similarly Br$\gamma$) is likely nebular. 
In addition, there is non--uniform nebular emission in the long slit image
of W43~\#3. Weak \ion{He}{2} 2.189~\mic \ absorption may be present in this
spectrum (EW \aple 1.5 \AA) 
but the S/N in the region around 2.189 \mic \ is insufficient for
an accurate measurement (the feature to the red side of 2.189 \mic \ is due
to incomplete telluric correction).

\subsection{Extinction to Individual Stars}

Rough estimates for the total line--of--sight extinction (interstellar plus 
circumstellar) were given in \S3.1.1. 
\ak \ and dereddened magnitudes can be derived for the spectroscopic
targets using an adopted extinction law (Mathis 1990\markcite{m90}), 
the observed colors
and the intrinsic colors. The latter can be estimated from the spectroscopic
classifications derived in the last sections. For O stars, the intrinsic colors
are small and the derived \ak \ depends most heavily on the extinction law
and observed colors. For W43~\#2 and \#3, we adopt the colors for an early
O star from Koornneef (1983\markcite{k83}) cited above in \S3.1.1.

\begin{figure}
\plotfiddle{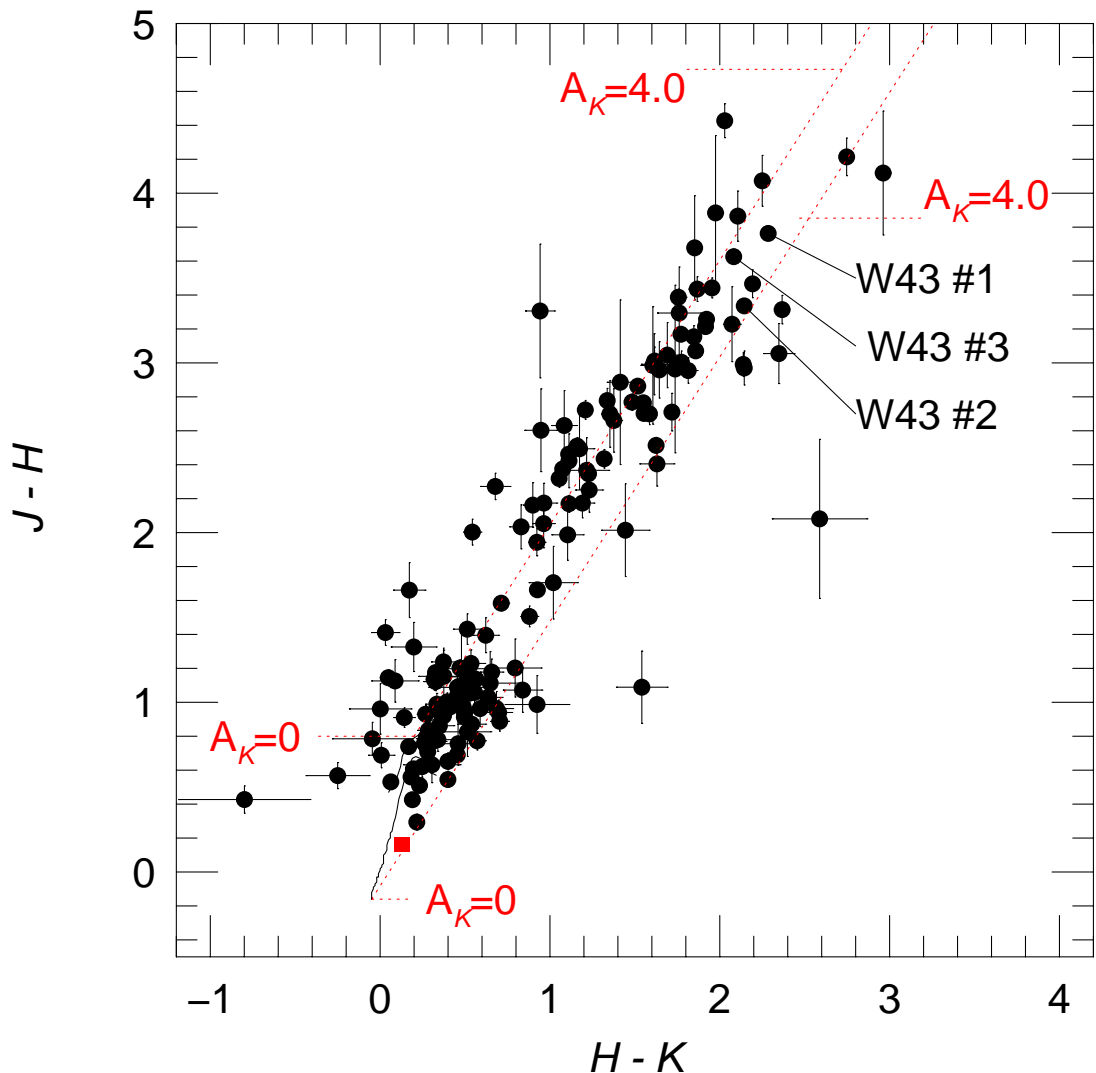}{6.0 in}{0}{80}{80}{-260}{-90}
\figcaption[]{$J-H$ vs. $H-K$ color--color diagram for the region
around the W43 cluster.
W43~\#1, \#2, and \#3, for which spectra are presented, are labeled.
No strong color excess is indicated for W43~\#1, \#2, or \#3 by comparison
with the reddening lines for early O stars ($H-K$ $=$ $-$0.05, 
$J-H$ $=$ $-$0.16) or M giants ($H-K$ $=$ 0.2, $J-H$ $=$ 0.8).
The color--color relation for un--reddened main sequence stars (Koornneef 1983)
and late--type giants (Frogel et al. 1978) is also shown for comparison.
The {\it square} point is the derived color for a WN7 star; see text.
\label{cc}
}
\end{figure}

For WN stars, the available data are sparse. We derive intrinsic $J-H$ and
$H-K$ using the the power law spectral indices of Morris et al. 
(1993)\markcite{meal93}. These models, derived from the observed spectral
energy distributions of WR stars, characterize the emitted flux by F$_\lambda$
$\sim$ $\lambda^{-\alpha}$. For the WN7 stars analized by  Morris et al.,
we find $\alpha =$ 3.11 $\pm$ 0.39. 
This corresponds to $J-H$ and $H-K$ $=$ 0.16 and 0.13, respectively
(this point is plotted in Figure~\ref{cc}). The
uncertainty in $\alpha$ leads to an uncertainty in each color index of
$<$ 0.15 mag. This, in turn, leads to an uncertainty in \ak \ of
$<$ 0.25 mag.

The positions of the three spectroscopic targets in Figure~\ref{cc}
are in good agreement with these adopted intrinsic colors. W43~\#3
lies on the reddening line for an early O star, W43~\#1 is close
to a projected line from the colors we derive for a WN7 star, and
W43~\#3 is at least within the envelope for normal stars.

Using these adopted colors, we find \ak \ $=$ 3.55 mag, 3.52 mag, and
3.63 mag for W43~\#1, \#2, and \#3, respectively. We have averaged the
derived extinction calculated using the $J-H$ and $H-K$ colors separately,
rather than attempting to deredden along lines parallel to 
those in Figure~\ref{cc}. For typical variations in the adopted extinction
law (Mathis 1990\markcite{m90}), the \ak \ will change by \aple 0.25 mag.
This uncertainty dominates over the photometric errors.

\section{DISCUSSION}

\subsection{Similarities to Known Starburst Clusters}

The dense stellar cluster at the heart of W43 with associated evolved
massive stars is reminiscent of the well studied clusters in NGC~3603
(Drissen et al. 1995\markcite{dmws95}) in the Galaxy and R136 in the
giant H~II region 30~Doradus in the Large Magellanic Cloud (LMC; Massey \&
Hunter 1998\markcite{mh98}). A common thread between these two starburst
clusters and W43 is the close spatial association of apparently evolved WN
stars and younger O3 stars. Our infrared spectral types are not precise
enough to assign detailed sub--types for W43~\#2 and W43~\#3, but the latter
is likely an early O star as discussed in \S3.2.2. Both Drissen et al. 
(1995\markcite{dmws95}) and Massey \& Hunter (1998\markcite{mh98}) conclude
that the WN objects are probably core hydrogen burning stars with such
intense mass--loss winds that their spectra resemble the more evolved WR
stars (see also de Koter et al. 1997\markcite{dhh97}).

The recently
discovered starburst cluster 20 pc in projection from the Galactic center,
the Arches cluster, is also similar to R136, NGC~3603, and W43 in
that it contains many (perhaps 100) O stars (Serabyn et al.
1998\markcite{ssf98}) as well as WN stars (Cotera et al. 1996
\markcite{cot96}) in a remarkably small volume. Like W43, the Arches
cluster is heavily obscured by interstellar extinction, and its stellar
content can only be studied at near infrared wavelengths. Nevertheless,
W43 and the Arches cluster suggest that nearby examples of the starburst
phenomena are not as rare as previously thought.

\subsection{Comments on Distance and Luminosity}

In principle, the derived dereddened colors and spectral type can be used
to estimate the distance to W~43. In practice, a meaningful estimate will
require more spectra of stars in the cluster and additional 
luminosity indicators in the near infrared spectroscopic classification 
schemes. We plan to work on both of these aspects in the future.
To illustrate the present difficulties, we consider the following.

W43~\#1 has a spectral type with implicit uncertainties given the
fact that, in at least one scenario (\S 3.2.1), 
it suggests a potential companion. Nevertheless, using an estimate from
Vacca \& Torres-Dodgen (1990)\markcite{vt90} for the $M_V$ 
($-$6.54 $\pm$ 0.43 mag) of a typical
WN7 (including those with $+abs$ classifications) and a derived
$V-K$ (0.12 $\pm$ 0.59) from Morris et al. (1993)\markcite{meal93}
a lower limit to the distance can be derived: 2800$^{+1400}_{-900}$ pc.
Where the uncertainty is dominated by the intrinsic scatter in $M_V$
and systematic uncertainty in 
$V-K$ due to the range of spectral index. 
This near distance compared to the radio distance
(7 kpc; see \S1) suggests W43~\#1 likely does have some contamination 
in the $K-$band brightness. Similarly, the average distance for the  
two O stars is: 4300$^{+1100}_{-900}$ pc
using $M_V$ ($-$6.44 $\pm$ 0.67 mag) from
Vacca et al. 1996\markcite{vgs96} and $V-K$ ($-0.93$) from 
Koornneef (1983)\markcite{k83}. If W43~\#2 and \#3 are taken to be giants
instead of supergiants, then the distance estimate would be $\sim$ 5700 pc. 

The above discussion can be reversed and the luminosity derived instead of the 
distance. Adopting the radio distance of 7 kpc (see \S 1) and
assuming all the objects are single stars, the absolute
magnitudes for W43~\#1, \#2, and \#3 are $M_V = -8.51, -7.69, -7.28$ mag, 
respectively. These can be compared to the derived 
$M_V$ for the most luminous O
stars (Vacca et al. 1996\markcite{vgs96}) and
WN7 stars (Vacca \& Torres-Dodgen 1990)\markcite{vt90}: 
$-6.4$, $-6.1$, and
$-5.78$, for O supergiants, giants, and dwarfs, respectively and
$-6.54$ for WN7 stars. 
Individual O star 
spectral types exhibit mean $M_V$ as luminous as $-7.4$ mag, and 
a typical standard deviation in the distribution of $M_V$ at any spectral type
is $\sim$ 0.6 mag
(Vacca et al. 1996\markcite{vgs96}). 
The standard deviation in the distribution of $M_V$
for WN7 stars (Vacca \& Torres-Dodgen 1990)\markcite{vt90} is 0.43 mag and
the most luminous object (WR~89) has $M_V$ $=$ $-7.3$ mag. 
At seven kpc, these numbers suggest the W43 objects (in particular, 
W43~\#1) may be extremely 
luminous, as has been suggested for evolved massive stars in the
center of the Galaxy (e.g., Tamblyn et al. 1996\markcite{t96} 
and Figer et al. 1998\markcite{f98}). For example, W43~\#1 has an estimated
$M_{\rm bol}$ of $-$11.5 mag compared to $M_{\rm bol}$ $\leq$ $-$11.7 mag for
the Pistol star in the Galactic center Quintuplet cluster 
(Figer et al. 1998\markcite{f98}). The $M_{\rm bol}$ for W43~\#1 is derived
using a bolometric correction
of $-$3.0 mag (Crowther et al. 1995\markcite{chs95}).
Obviously, no definitive statements
can yet be made for W43 with only three spectra in hand and in the
face of the substantial uncertainties outlined here.

Intrinsic colors and magnitudes for rare spectral types
like W43~\#1 may always provide difficulties. 
However, the colors for more normal O and B stars are well
understood. As our spectral classification schemes
improve and we obtain spectra of fainter O and B stars
in the cluster, we will be able to better 
assess the distance to W43 and the luminosity of the individual stars. 
It is already clear that clusters like W43
will be very important in comparison to the Galactic center star
clusters
and to the further study of massive stars in the Milky Way.

\subsection{Star Formation}

The combination of the Lester et al. (1985)\markcite{ldw85} far infrared
maps and the present near infrared images and spectra suggest the 
following scenario. The stellar cluster appears between the strongest far
infrared sources (G30.08 N and S) in a region of lower dust emission.
The evolved nature of the central massive stars in the cluster and lower
dust emission indicate the central cluster may have formed first and 
subsequently cleared away the remains of
the gas and dust which came together to form it
(or is in the process of clearing it away: there are dark regions
very close in projection to the central cluster). The embedded red objects 
seen in Figure~\ref{clr2} in the regions coincident with the far infrared
sources G30.08 N and S may be newly forming stars more deeply embedded in
the molecular clouds now being ionized by the central cluster (and to 
some extent by massive stars forming within).
We will test this scenario by searching for spectroscopic signatures
of YSOs in the brighter embedded objects in G30.08 N and S.

\section{SUMMARY}

Near infrared images reveal a dense cluster of newly formed stars at the
core of the Galactic giant H~II region W43. Spectra of three of the brightest 
cluster members show them to be hot, massive stars. The brightest member,
W43~\#1, has a $K-$band spectrum very similar (in morphology as well as line
strength and width) to similar spectra for optically classified WR stars
of the WN7 type. Recent observations near the Galactic center, in NGC~3603,
and in R136 in the LMC all find WN type stars at the center of dense 
starburst clusters.
A picture is emerging where these objects are thought to be 
core burning H stars with extreme mass--loss, rather than the somewhat
older evolved counterparts of initially massive O stars. 

The combination of $J$, $H$, and $K$ images reveals a complex morphology
in the region surrounding the stellar cluster. In particular, the spatial
geometry suggests the starburst cluster may have triggered second generations
of stars to form in the molecular material to the North and South.
Both these regions are locations of far infrared emission maxima which suggest
even younger stars still more highly obscured.

It is a pleasure to acknowledge the help of J. Holtzman in installing the 
latest version of LickVista (5.0) on RDB's new Linux box. We appreciate 
the availability and support of quality tools like LickVista. We also thank
R. Pogge for his Liner program used to analize spectra presented here.
We kindly acknowledge suggestions from 
an anonymous referee which have improved our paper.
PSC appreciates support by the NSF under grant 97-31520.
A.D. thanks PRONEX/FINEP for financial support.
Support for this work was provided by NASA through grant number
HF 01067.01 -- 94A from the Space Telescope Science Institute, which is
operated by the Association of Universities for Research in Astronomy,
Inc., under NASA contract NAS5--26555.

\newpage


\newpage


\begin{deluxetable}{lcc}
\tablecaption{W43~\#1 $K-$band Spectra Data\label{tab1}}
\tablehead{
\colhead{Line Id (\mic)} &
\colhead{FWHM (\AA)} &
\colhead{EW (\AA)}
}
\startdata
 2.0378 \ion{He}{2} &   144 &5  \nl
 2.1127 \ion{He}{1}, 2.1155 \ion{N}{3} &   103 &11  \nl
 2.1661 \ion{H}{1}&   160 &48  \nl
 2.1892 \ion{He}{2} &   109 & 11
\enddata
\end{deluxetable}

\end{document}